\documentclass[%
 reprint,
 amsmath,amssymb,
 aps,
]{revtex4-1}

\usepackage{graphicx}
\usepackage{dcolumn}
\usepackage{bm}

\begin{document}


\title{Observation of a diffractive contribution to dijet production in proton-proton collisions at $\sqrt{s}=7$~TeV}%

\author{Diego Figueiredo}
\affiliation{Physics Department, Rio de Janeiro State University}
\email{dmf@cern.ch}
\collaboration{on behalf of the CMS Collaboration}


\date{\today}

\begin{abstract}
The differential cross section measurement for dijet production at proton-proton collisions at $\sqrt{s}=7$~TeV is presented as a function of an approximation for the fractional momentum loss of the scattered proton ($\tilde{\xi}$), an useful observable to distinguish model predictions of diffractive and nondiffractive components. The data was collected by the CMS detector at low instantaneous luminosity runs at LHC, corresponding to an integrated luminosity of $2.7~{\rm nb}^{-1}$. This is the first observation of single diffractive dijets at LHC\cite{dijets}.
\end{abstract}

\maketitle


\section{\label{sec:level1} Overview}

Particle physics diffractive phenomena is experimentally correlated with the presence of large rapidity gaps (LRGs), or regions devoid of activity in pseudorapidity ($\eta$) space. These types of processes are explained by strongly interacting color-singlet exchanges with vacuum quantum numbers or the so called Pomeron ($\mathbb{P}$) trajectory. Due to the exchange with the vacuum quantum numbers, and specially no color, there is no hadronic activity in the large rapidity range adjacent to the scattered proton (or protons)\citep{collins}.

Diffractive events with a hard parton-parton scattering are interesting since they can be studied in terms of perturbative quantum chromodynamics (pQCD). In this framework, the proton-proton ($pp$) diffractive cross section is suppressed due to the effect of soft rescattering between the two protons, known as gap survival probabilty. Such suppresion is related to the breaking of the hard scattering factorization in diffractive processes \cite{diffractiondidatic}. 

\begin{figure}[htb]
\includegraphics[scale=0.4]{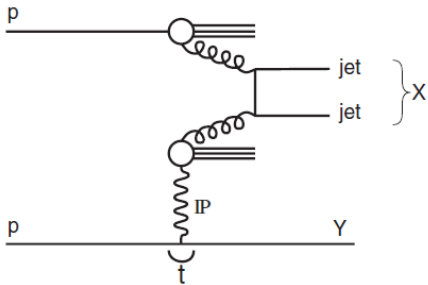}
\caption{\label{fig:topology} Event topology of single diffractive dijet production. While a proton is broken up produzing two jets after a color-singlet exchange, the scattered proton remains intact with almost the same initial momentum.}
\end{figure}

The observation of single diffractive dijet production with the CMS detector is presented. Fig.~\ref{fig:topology} shows the event topology of such events. Experimentally, it is a clean signature characterized by one proton (or an excited state $Y$) with low momentum loss ($\xi$) and two high $p_{T}$ jets. Events with LRGs are detected. 

The observed cross section is presented as a function of a variable that approximates the scattered proton momentum loss, and is related to the dissociative diffractive mass ${\rm M_x}$. It is defined, assuming that the proton is scattered in the positive ($\tilde{\xi}^{+}$) or negative ($\tilde{\xi}^{-}$) CMS side, as:

\begin{equation}
 \tilde{\xi}^{\pm}=\frac{\sum({\rm E}^{\rm i} \pm {\rm p}_{\rm z}^{\rm i})}{\sqrt{s}},
\end{equation}

where ${\rm E}^{\rm i}$ and ${\rm p}_{\rm z}^{\rm i}$ are the energy and longitudinal momentum of the ith final-state particle with $-\infty < \eta < 4.9$ for $\tilde{\xi}^{+}$ and $-4.9 < \eta < +\infty$ for $\tilde{\xi}^{-}$. In the region of low $\tilde{\xi}^{\pm}$, this variable is a good approximation of $\xi$ for single diffractive events.

\section{\label{sec:level2}Experimental Setup}

\begin{figure}[htb]
\includegraphics[scale=0.24]{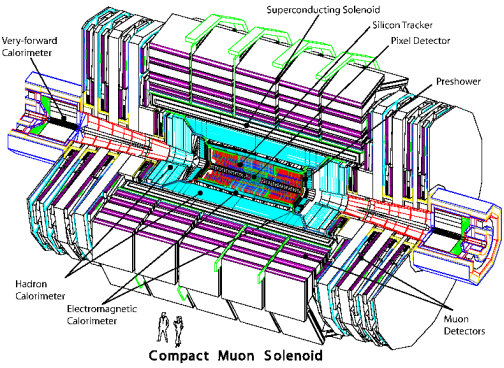}
\caption{\label{fig:detector}The CMS Detector.}
\end{figure}

The CMS detector \cite{cms}, Fig.~\ref{fig:detector}, has a superconducting solenoid able to generate a magnetic field of $3.8$~T. The solenoid surrounds the tracker system, with acceptance of $\vert\eta\vert < 2.5$, composed of pixel detectors and silicon strips, the electromagnetic calorimeter (ECAL), $\vert\eta\vert < 3$, composed of crystals of lead-tungstate and hadronic calorimeter (HCAL), $\vert\eta\vert < 3$, composed of brass and scintillators.

The ECAL is a segmented calorimeter with transverse dimensions $\Delta\eta\times\Delta\phi=0.0174\times 0.0174$, where $\phi$ is the azimuthal angle in radians. The cells of the HCAL towers are grouped into granularity $\Delta\eta\times\Delta\phi = 0.087\times 0.087$  in the central region of rapidity and increasing gradually to reach the forward region of detector. The forward calorimeter, named HF and with acceptance of $3.0<\vert\eta\vert < 5.0$, being composed of quartz scintillators fibers and photomultipliers is used in this analysis to detect LRGs. Muons are measured in gas chambers detectors, with acceptance $\vert\eta\vert < 2.4$, and are surrounded by iron plates. The polar angle $\theta$ is measured in the anticlockwise proton beam direction. The pseudorapidity $\eta$ is defined as $-\ln[\tan(\frac{\theta}{2})]$ and is equal to the rapidity $Y$ in the limit of a massless particle.

\section{\label{sec:level3}Event Selection}

The following criteria were used to select events consistent with single diffractive dijet production: 

\begin{enumerate}
\item At least one jet with uncorrected $p_T>6$~GeV was required to readout the CMS detector. The trigger efficiency is estimated to be around $95\%$; 
\item off-line, events were selected with at least two jets, reconstructed by the anti-${\rm k}_{\rm t}$ jet finding algorithm with distance parameter of $0.5$ using particles reconstructed with a Particle Flow (PF) algorithm~\citep{tdrsoft}, and corrected to particle level. Both jets were required to have $p_T>20$~GeV;
\item to enhance the single diffractive contribution, the requirements $\eta_{\rm max} < 3$ or $\eta_{\rm min} > -3$ were also applied. Here $\eta_{\rm max}$ ($\eta_{\rm min}$) is the pseudorapidity of the most forward or backward PF object per event. Events selected with this cut have no activity in the forward or backward region of the detector.
\end{enumerate}

\begin{figure}[htb]
\includegraphics[scale=0.4]{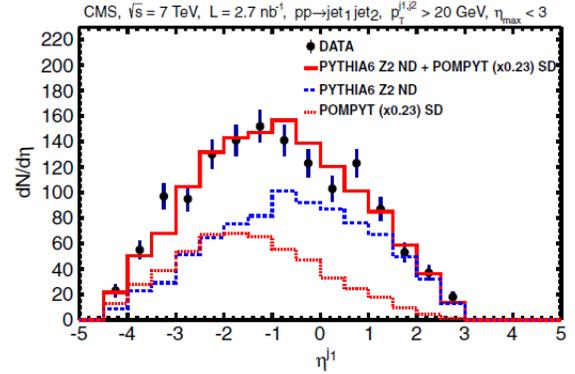}
\caption{\label{fig:etajet1}Reconstructed pseudorapidity distributions of the leading jet after the $\eta_{\rm max} < 3$ selection (black dots) compared to three detector-level MC simulations (histograms). Events with the $\eta_{\rm min} > -3$ condition are also included in the figure with $\eta^{\rm j1,j2}\rightarrow-\eta^{\rm j1,j2}$. The error bars indicate the statistical uncertainty. The sum of the predictions of the two MC simulations is normalized to the number of events in the corresponding distributions for the data.}
\end{figure}

The leading jet $\eta$ distribution, Fig.~\ref{fig:etajet1}, shows that the shape of PYTHIA6\citep{pythia6} (non-diffractive), in red solid line, and POMPYT\citep{pompyt} (diffractive) Monte Carlo (MC) fits the data, after all selection criteria are applied.

\begin{figure}[htb]
\includegraphics[scale=0.48]{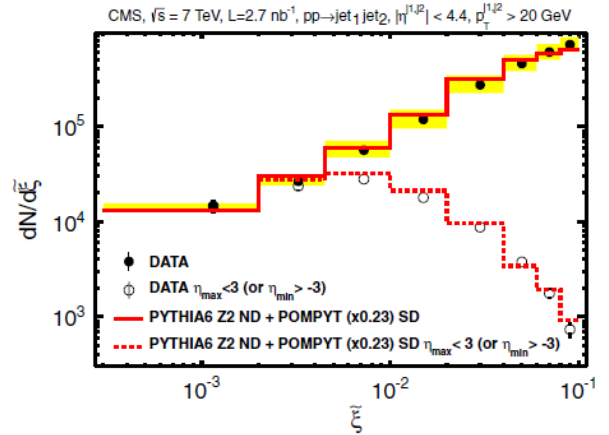}
\caption{\label{fig:cross}Reconstructed $\tilde{\xi}$ distributions with (open symbols) and without (closed symbols) the $\eta_{\rm max} < 3$ (or $\eta_{\rm min} > -3$) conditions are compared to detector-level MC predictions. The error bars indicate the statistical uncertainty; the band represents the calorimeter energy scale uncertainty. The sum of the predictions of the two MC simulations is normalized to the number of events in the corresponding distributions for the data.}
\end{figure}

Fig.~\ref{fig:cross} shows the reconstructed $\tilde{\xi}$ distributions with and without the $\eta_{\rm max}$ requirements. According to Fig.~\ref{fig:etajet1} and \ref{fig:cross}, a diffractive contribution is seen in the data. The data is compared with diffractive and non-diffractive MC predictions as a differential cross section as a function of $\tilde{\xi}$.

\section{\label{sec:level4}Results}

The differential cross section, ${\rm d}\sigma_{\rm jj}$/${\rm d}\tilde{\xi}$, is defined as:

\begin{equation}
\frac{{\rm d}\sigma_{\rm jj}}{{\rm d}\tilde{\xi}} = \frac{{\rm N}^{\rm i}_{\rm jj}}{{\rm L}\cdot\epsilon\cdot{\rm A}^{\rm i}\cdot\Delta\tilde{\xi}^{\rm i}}
\end{equation}

where ${\rm N}^{\rm i}_{\rm jj}$ is the measured number of dijet events in the ith $\tilde{\xi}$ bin, ${\rm A}^{\rm i}$ is the correction factor defined as the number of reconstructed MC events in that bin divided by the number of generated events in the same bin, $\Delta\tilde{\xi}^{\rm i}$ is the bin width, {\rm L} is the integrated luminosity, and $\epsilon$ is the trigger efficiency. The factors ${\rm A}^{\rm i}$ include the effects of the geometrical acceptance of the detector, and of all event selection criteria as the unfolding corrections due to the finite resolution of the reconstruction.

\begin{figure}[htb]
\includegraphics[scale=0.45]{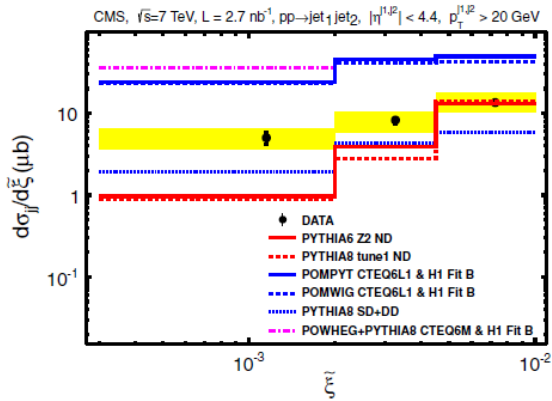}
\caption{\label{fig:crossxi}Differential cross section for inclusive dijet production as a function of $\tilde{\xi}$. The points are plotted at the center of the bins. The error bars indicate the statistical uncertainty and the band represents the systematic uncertainties added in quadrature.}
\end{figure}

\newpage

The corresponding values for each bin, Fig.~\ref{fig:crossxi}, are shown in Table~\ref{tab:table1}.

\begin{table}
\caption{\label{tab:table1}Cross section for inclusive dijet production as a function of $\tilde{\xi}$ for jets with $p_{T}^{{\rm j1,j2}}>20$~GeV and jet axes in the pseudorapidity range $\vert\eta^{{\rm j1,j2}}\vert < 4.4$.}
\begin{ruledtabular}
\begin{tabular}{lcr}
{\bf $\tilde{\xi}$ Bin} & {\bf $\frac{{\rm d}\sigma_{\rm jj}}{{\rm d}\tilde{\xi}}$} ($\mu$b) \\
\hline
$0.0003 < \tilde{\xi} < 0.002$ &  $5.0\pm 0.9{\rm (stat)}^{+1.5}_{-1.3}{\rm (sys)}$ \\
$0.002 < \tilde{\xi} < 0.0045$ &  $8.2\pm 0.9{\rm (stat)}^{+2.2}_{-2.4}{\rm (sys)}$ \\
$0.0045 < \tilde{\xi} < 0.01$  &  $13.5\pm 0.9{\rm (stat)}^{+4.5}_{-3.1}{\rm (sys)}$ \\ 
\end{tabular}
\end{ruledtabular}
\end{table}

\section{\label{sec:level5}Summary}

The dijet differential cross section as a function of the variable $\tilde{\xi}$ has been presented. It is the first observation of single diffractive dijet production at LHC. Based on this measurement, it is possible to estimate the effect of the suppression of the diffractive cross section in $pp$ collisions. Leading-order (LO) diffractive generators (POMPYT or POMWIG\citep{pomwig}), based on dPDFs from the HERA experiments\cite{h1fit}, overestimate the measured cross section, as seen in Fig.~\ref{fig:crossxi}. In the first $\tilde{\xi}$ bin, their normalization needs to be scaled down by a factor of $\approx 5$. This factor can be interpreted as the effect of the gap survival probability. The results are also compared to NLO predictions (POWHEG~\citep{powheg}).

Taking into account the contribution of proton-dissociative events, based on a MC prediction, the estimated gap survival probability based on POWHEG and POMPYT (or POMWIG) are shown in Table~\ref{tab:table2}.

\begin{table}
\caption{\label{tab:table2}Gap survival probability estimated in the range $0.0003 < \tilde{\xi} < 0.002$ for NLO and LO diffractive generators.}
\begin{ruledtabular}
\begin{tabular}{lcr}
{\bf $\langle{\rm S}^2\rangle^{\rm MC}$ ($0.0003 < \tilde{\xi} < 0.002$)} & {\bf Monte Carlo} \\
\hline
$0.08 \pm 0.04$ &  POWHEG (NLO) \\  
$0.12 \pm 0.05$ &  POMPYT or POMWIG (LO) \\
\end{tabular}
\end{ruledtabular}
\end{table}

\begin{acknowledgments}
I would like to thank for FAPERJ to support and for LISHEP 2013 committee and editors.
\end{acknowledgments}


\bibliography{diffraction}

\end{document}